\documentclass{iopart} 
\usepackage{iopams} 
\usepackage{url}
\usepackage{hyperref}
\hypersetup{
    colorlinks,
    citecolor=black,
    filecolor=black,
    linkcolor=black,
    urlcolor=black
}
\usepackage{color}

\usepackage{graphicx}
\newcommand{\aligo}{Advanced LIGO}
\begin{document}
\renewcommand{\thefootnote}{\arabic{footnote}}

\title{Sensors and Actuators for the Advanced LIGO Mirror Suspensions}

\author{L.\,Carbone$^1$, 
S.M.\,Aston$^1$, 
R.M.\,Cutler$^1$, 
A.\,Freise$^1$,\\
J.\,Greenhalgh$^2$, 
J.\,Heefner$^3$, 
D.\,Hoyland$^1$, 
N.A.\,Lockerbie$^4$,\\
D.\,Lodhia$^1$, 
N.A.\,Robertson$^{3,5}$, 
C.C.\,Speake$^1$, 
K.A.\,Strain$^5$, 
and A.\,Vecchio$^1$ 
}

\address{$^1$ School of Physics and Astronomy, University of Birmingham, Edgbaston - 
Birmingham, B15\,2TT, UK \\
$^2$ 
Science and Technology Facilities Council, Rutherford Appleton Laboratory, Didcot, Oxon, OX11 OQX, UK\\
$^3$ LIGO Laboratory, California Institute of Technology, MS\,100-36, 
Pasadena, CA 91125, USA\\
$^4$ 
Department of Physics, University of Strathclyde, SUPA, John Anderson Building, 107 Rottenrow, Glasgow, G4 0NG, UK\\
$^5$ 
SUPA, Institute for Gravitational Research, School of Physics and Astronomy,
University of Glasgow, Glasgow, G12 8QQ, UK }

\ead{lc@star.sr.bham.ac.uk}

% Classical and Quantum Gravity 29 115005 (2012)
% http://iopscience.iop.org/0264-9381/29/11/115005/

\begin{abstract}
We have developed, produced and characterised integrated sensors, 
actuators and the related read-out and drive electronics that will be used for the control of the \aligo~suspensions. 
The overall system consists of the BOSEMs (displacement sensor with integrated electro-magnetic actuator), the satellite boxes (BOSEM readout and interface electronics) and six different types of coil-driver units. 
In this paper we present the design of this read-out and control system, we discuss the related performance relevant for the Advanced LIGO suspensions, and we report on the experimental activity finalised at the production of the instruments for the \aligo~detectors. 
\end{abstract}

\section{Introduction}
The first generation of ground-based gravitational wave interferometers like LIGO, GEO\,600 and VIRGO \cite{LIGO,GEO,VIRGO} 
have reached their design sensitivities and are currently in the process of being upgraded to second generation - `advanced' - facilities \cite{harry:CQG:2009,geohf,avirgo}. 
The planned modifications aim to 
widen the sensitivity band of the observatories and increase their detection range: 
in the particular case of 
\aligo~\cite{harry:CQG:2009}, the upgrade will improve the detector's sensitivity by about an order of magnitude over almost all the sensitive band, for a best strain sensitivity goal 
of about $3 \times 10^{-24}$\,Hz$^{-1/2}$ 
around few hundred Hz, 
and an explorable region of the universe thousand 
times larger. 

Among the different core subsystems of the interferometer 
undergoing a major upgrade in \aligo, a crucial role will be played by the combined effect of an enhanced seismic isolation system \cite{ligo:seis} 
together with a novel
multi-stage mirror suspension design \cite{robertson:CQG:2002}. 
This will provide an improved isolation
from seismic disturbances by at least three orders of magnitude in the 1\,Hz to 10\,Hz region. The reduced 
residual motion of the test-masses - approximately $10^{-19}$\,m\,Hz$^{-1/2}$ 
at 10\,Hz -
will extend the useful window for detection of gravitational wave signals 
from the original 
$\sim 40$\,Hz
down to 
10\,Hz, making 
background Newtonian noise and
Brownian noise of the last stage 
of the mirror suspensions  
the ultimate noise sources 
at low frequencies \cite{ref:monolithics}. 
\begin{figure}[t]
\begin{center}
\includegraphics[width=12.8cm]{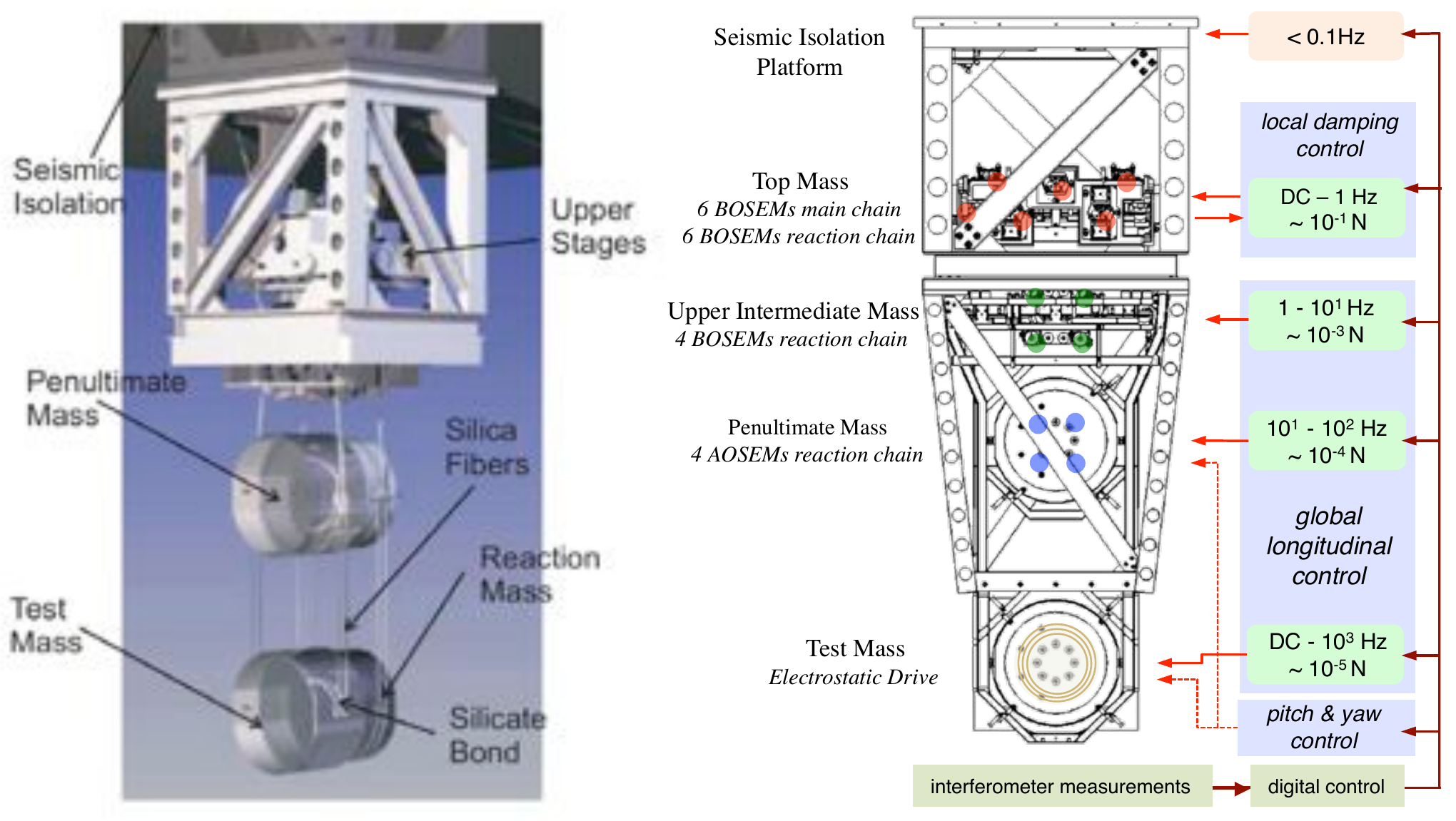}
\caption{\label{fig:quad} 
The \aligo~quadruple suspension systems: Left, 3D CAD representation of the suspensions \cite{harry:CQG:2009}; Right, location of sensors and actuators within the quadruple suspensions for the top (red dots), upper-intermediate (green) and penultimate stages (blue). 
A sketch of the control loop scheme, with main frequency range of operation for each actuation stage and related maximum actuation force range are also shown.
}
\end{center}
\end{figure}

A picture of the quadruple pendulum suspension system, which is used for the isolation of the arm cavities test-masses,
is shown in Fig.\,\ref{fig:quad}\,(left). 
Here the key elements are 
the three stages of cantilever spring-blades, which form the upper-half of the suspension chain and
which provide enhanced vertical-motion seismic isolation, and
the 40\,kg fused silica 
test-mass, monolithically suspended from the `twin' penultimate mass by means of silicate-bonded fused silica fibres, for a ultra-low suspension thermal noise. 
The four-stage pendulum hangs from the `support structure', which connects it to the seismic isolation platform and which is used to host the sensors and actuators used for the damping of the suspension resonances. 
Alongside the main suspension chain, `reaction-masses' are independently suspended in parallel 
to the main chain and provide a seismically quiet platform for mounting the sensors and the actuators 
that, in combination with the ones 
acting from the support structure, 
allow for control of 
the arm-lengths of the interferometer.

The \aligo-UK collaboration (Science and Technology Facilities Council, 
University of Glasgow, University of Birmingham and University of Strathclyde)
has developed the concept and the design, and
has played a leading role in providing the novel and improved technologies needed for the \aligo~suspension subsystem  \cite{ALUK:proposal}. 
Within this collaborative effort, 
the Birmingham group has 
been responsible for the design, prototyping and development of 
an integrated system of 
sensors, actuators and 
analogue electronics 
for the dynamic control of the suspensions, 
and for the subsequent large-scale production of such devices. 
A significant fraction of the production work has been in the testing, characterisation and validation
of each of the units, to ensure that the functional and scientific requirements of the instruments are met.
In this paper we 
overview the role of this sensing and actuation system
within the \aligo~suspensions, we present the design and performance of such instruments 
and we discuss the results of experiments undertaken at the University of Birmingham for their production and validation. 

\section{Sensors and Actuators for the \aligo~suspensions \label{Sec:2}}
The integrated system of sensors and actuators developed for the control of the 
\aligo~suspension stages consists of three types of instruments:
\begin{itemize}
\item BOSEMs (`Birmingham Optical Sensor and Electro-Magnetic actuator'), 
integrated optical position sensor and electro-magnetic actuator;

\item Satellite boxes, electronic interfaces for the BOSEM displacement sensor; 

\item Coil-driver units, electric current drivers for the BOSEM actuator. 
\end{itemize}

\begin{figure}[b]
\begin{center}
\includegraphics[width=13.0cm]{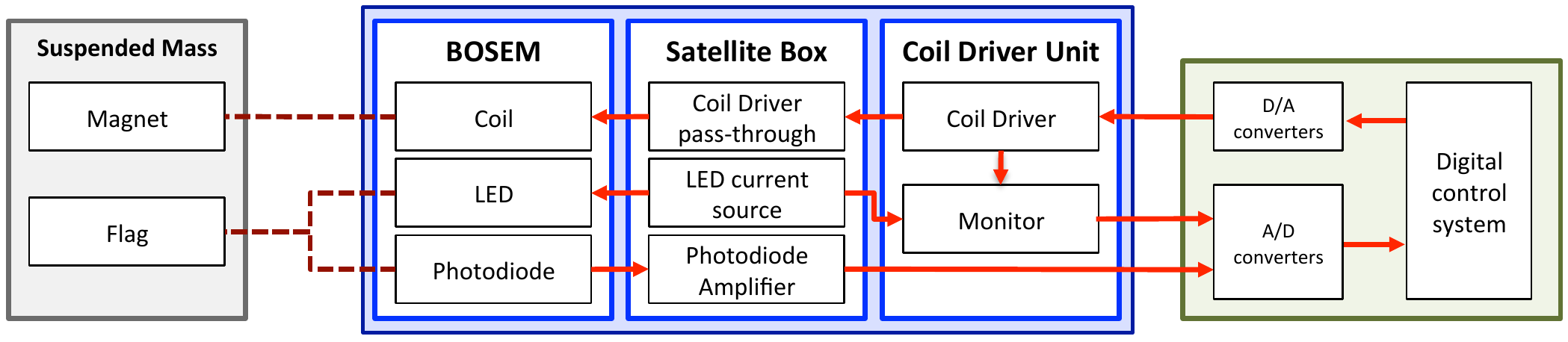}
\caption{\label{fig:schematic} Schematic of the working principle 
of the integrated system of BOSEMS, satellite boxes and coil-drivers. }
\end{center}
\end{figure}

\noindent The working principle of the integrated BOSEM read-out and actuation 
system is described with the diagram in Fig.\,\ref{fig:schematic}. 
The motion of the suspended mass is sensed by the BOSEM, whose signal is filtered and amplified by the satellite box, and sent to the main digital control system.
Here the signal is 
processed and purposely optimised
to be then fed back through the coil-driver amplifier to the BOSEM coil, to eventually actuating on the suspended masses. 
Within the \aligo~suspensions, this 
system is used to monitor and control the 
suspension stages during the initial alignment, damping and 
acquisition 
of the different interferometer subsystems, and subsequently to support, in combination with the interferometric measurements, the 
control of the position of the suspended optics to allow for quiet operation of the interferometer science mode.

\begin{figure}[ht]
\begin{center}
\includegraphics[width=12.3cm]{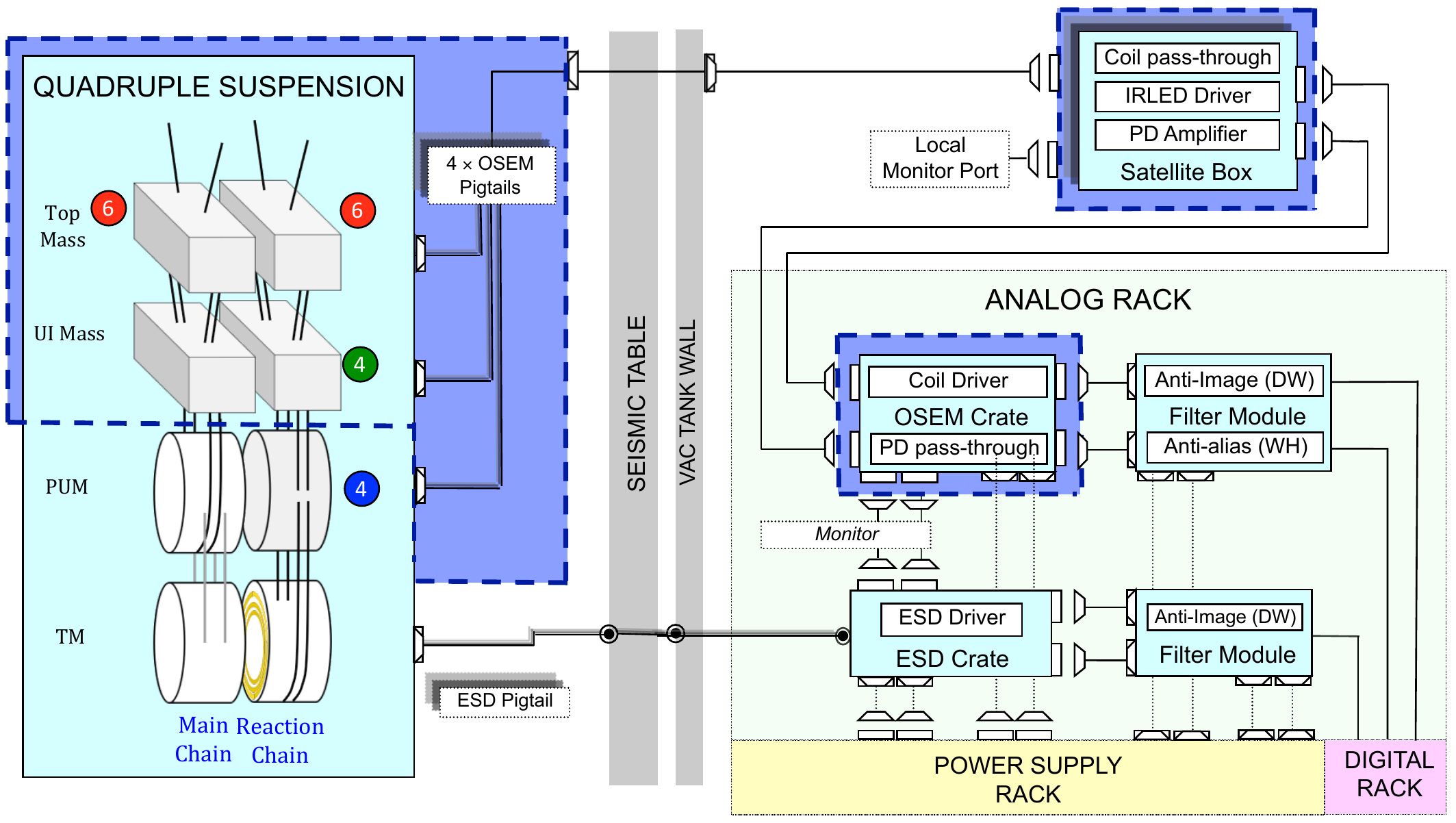}
\caption{\label{fig:wherewhat} Scheme of the role of BOSEMs and related electronics within the \aligo~quadruple suspension control scheme. 
The instruments described in this paper are highlighted within the dark-blue dashed areas. 
}
\end{center}
\end{figure}

A sketch of the \aligo~quadruple suspension's control loop scheme, 
showing the role of the instrumentation discussed in this paper, 
is given  in Fig.\,\ref{fig:quad}\,(right) 
and in Fig.\,\ref{fig:wherewhat}. 
The BOSEM's system provides optical sensing and electro-magnetic actuation capabilities in the three top stages.
Six units are used to provide sensing and control on all six d.o.f of the top-masses of the main suspension chain, and other six are similarly employed for the reaction chain;  
four BOSEMs units control the swing, pitch and yaw modes of the main chain second stage, acting from the reaction one; 
due to the reduced actuation forces required to actuate on the penultimate suspension stage, 
satellite boxes and coil-drivers are here integrated with a similar but compact and lightweight sensing and actuation instrument, the re-engineered version of Initial LIGO OSEMs \cite{fritschel:DCC:1999} now dubbed AOSEMs \cite{AOSEM}; 
finally the smaller amplitude, higher frequency actuation forces required to control the mirror test-masses, on the fourth and bottom stage, are provided electro-statically with a different type of instrument, named the Electrostatic Drive \cite{elec:drive}.

The integrated BOSEMs system allows sensing 
of the motion of the suspended masses 
with displacement sensitivities 
of the order of $3\times10^{-10}$\,m\,Hz$^{-1/2}$ or better
in the frequency range from 0.1\,Hz up to 10\,Hz and above, 
and provides at the same time strong actuation authority on the masses 
with overall force peaks
of order hundreds mN on each d.o.f. 
and damping times of order 10\,s combined with 
very low actuation force noise, at the pN\,Hz$^{-1/2}$ level and below. 
With the four pendulum stages naturally suppressing the residual longitudinal motion of the lowest 
mass by about seven orders of magnitude (from $\sim10$\,Hz), 
the BOSEMs sensitivity allows initial damping of the test mass motion 
down to a level of $\approx10^{-17}$\,m\,Hz$^{-1/2}$,
at frequencies around few Hz and above.
The seismic isolation goal, namely $10^{-19}$\,m\,Hz$^{-1/2}$ at 10\,Hz, is then achieved 
combining the BOSEM system with interferometer and cavity signals - with sensitivity well below the pm\,Hz$^{-1/2}$ level \cite{robertson:CQG:2002} - and 
with the electrostatic drive actuation - with residual forces at fN\,Hz$^{-1/2}$ level or better -
and by making use of sophisticated damping and control algorythms, see for example \cite{shapiro:2012,strain:2012}.

Compared to different typologies of displacement sensors 
used in other GW detectors \footnote[8]{It is important to note that GEO\,600 used OSEM-like sensors and actuators, 
and that the BOSEMs presented here are now in use in the GEO\,600 upgrade, GEO-HF.} 
(e.g., accelerometers, LVDTs, CCD-camera based optical levers \cite{virgoSA,virgoLS}),
the BOSEMs have been selected for the reduced dimensions and the higher 
sensitivity provided, other than of course for continuity with the Initial LIGO design.
At the same time, 
thanks to the relaxed sensing requirements,
the BOSEMs have been preferred to potentially more sensitive 
interferometric sensor designs, also investigated
during the preliminary study phase \cite{speake:2005}, for the more mature technology and proven reliability, and
for the known contained costs and relative simplicity of large scale manufacturing. 
Concerning the coil magnet actuation scheme, which is
widely used in all other GW detector suspensions, a crucial role in the selection of the
BOSEMs design was played by the compactness of the design itself, with sensing and actuation in the
same device, which made them the favourable choice for the simplicity of installation,
testing and eventually operation.
The BOSEMs and related electronics provide a similar sensing and actuation 
functionality to the `shorter' suspension types 
(`triple' and `double' suspensions), 
which are used for optical elements of the interferometer 
demanding less stringent seismic isolation, such as for instance beam-splitters and mode-cleaner optics. 
The BOSEMs are here used for 
the upper stages, the AOSEMs for the lower 
ones, and satellite boxes and coil-drivers serve the sensors and actuators at all stages. 
In total, about 
185 BOSEM actuation units are employed for local and global control of each \aligo~interferometer. 
The overall production run at Birmingham delivered to \aligo~consists 
of 650 BOSEMs, 230 satellite boxes, and 267 Coil Driver units 
for the three \aligo~interferometers. This includes about 25\% spare units produced for redundancy and replacement in the event of device failure, and about 100 units for the interferometer prototypes. 

\subsection{BOSEM \label{Sec:BOSEM}}
\begin{figure}[t]
\begin{center}
\includegraphics[width=13.5cm]{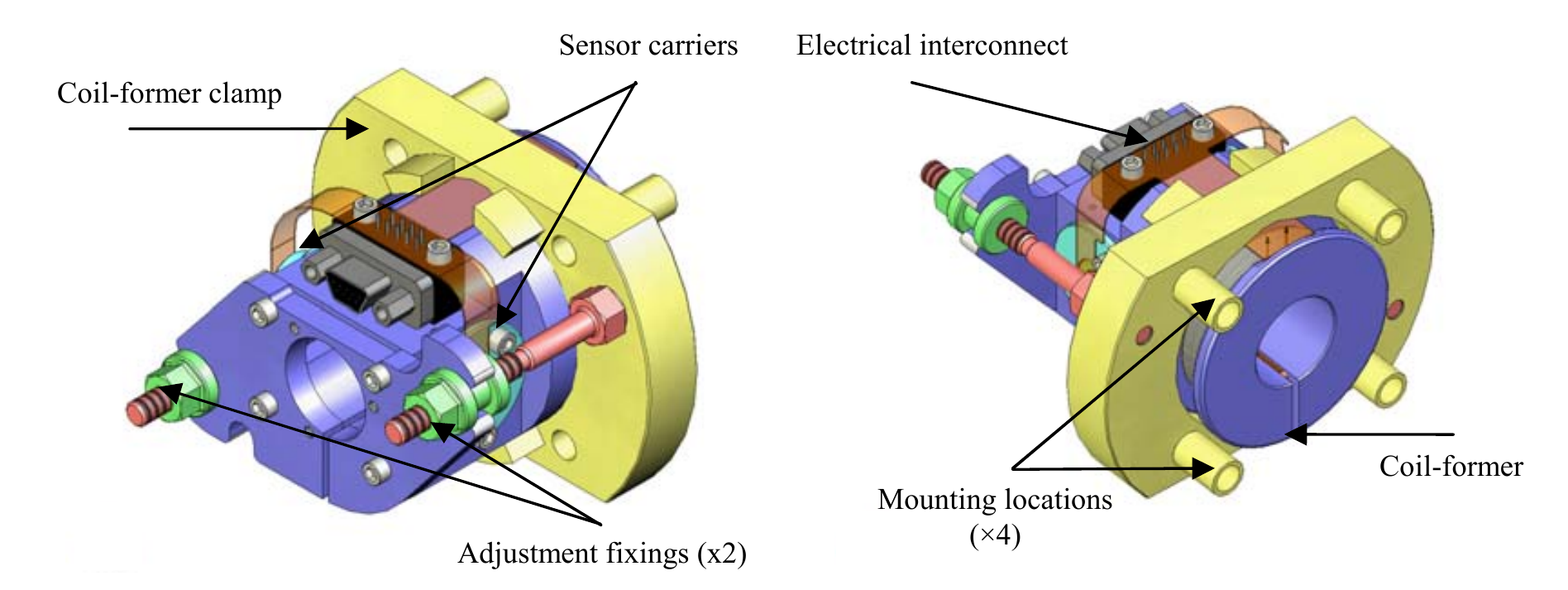}
\caption{\label{fig:BOSEM:des} Front and rear view CAD designs of the BOSEMs. }
\end{center}
\end{figure}

\begin{figure}[b]
\begin{center}
\includegraphics[height=3.3cm]{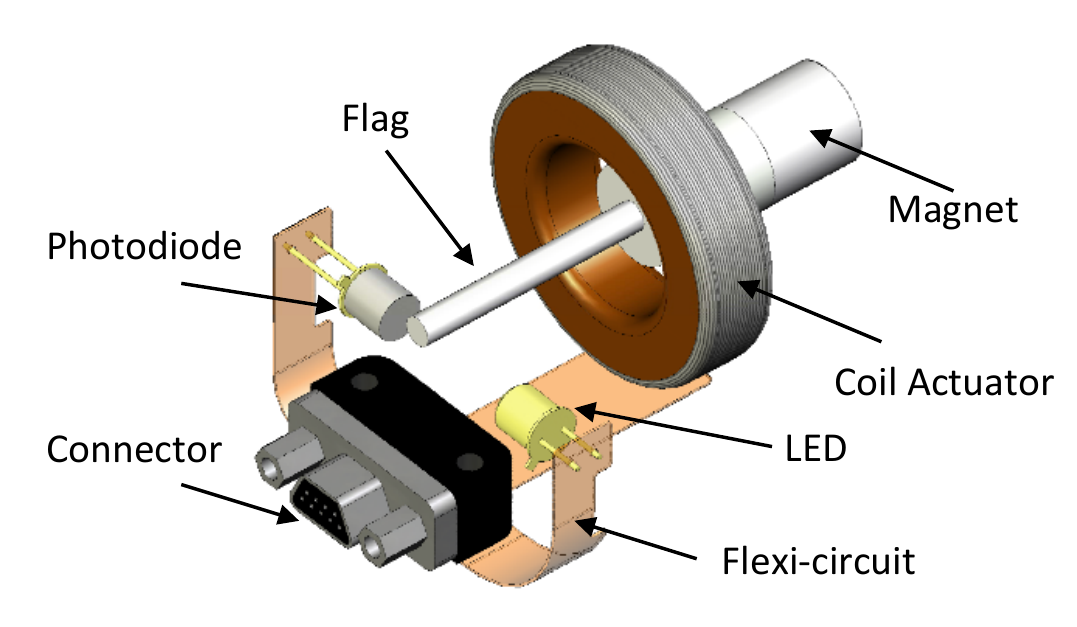}
\includegraphics[height=3.3cm]{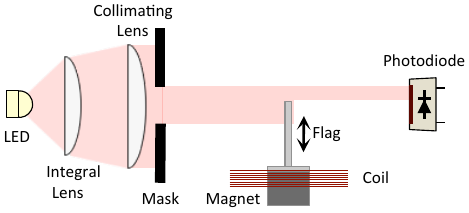}
\caption{\label{fig:BOSEM:scheme} Left: Schematic of the integrated BOSEMs subsystems. Right: cartoon showing the BOSEMs shadow sensing scheme. Relevant dimensions are the mask aperture, 1.5\,mm$\times$4.5\,mm, and the mask-PD separation, $5$\,mm. }
\end{center}
\end{figure}

The BOSEMs  
are compact, ultra-high-vacuum (UHV) compatible, 
non-contact,
low-noise 
position sensors with integrated electro-magnetic actuator. 
The BOSEMs working principle is based on the original concept by Shoemaker {\it et al} \cite{shoemaker},
and their final design 
builds upon a series of incremental 
modifications to the Initial LIGO OSEM and on the subsequent sensor study and optimisation process
described in \cite{romie:DCC:2003,strain:DCC:2004:feb,strain:DCC:2004,lockerbie:DCC:2004:1,lockerbie:DCC:2004:2}. 
Each unit comprises several interconnected subsystems:  optical read-out; coil-magnet actuator; magnet-flag assembly; alignment system; electrical connection. 
CAD models of the BOSEMs assembly are shown in Fig.\,\ref{fig:BOSEM:des}, while
a sketch of the integrated subsystems is given in Fig.\,\ref{fig:BOSEM:scheme}\,(left).

\begin{figure}[t]
\begin{center}
\includegraphics[width=13.4cm]{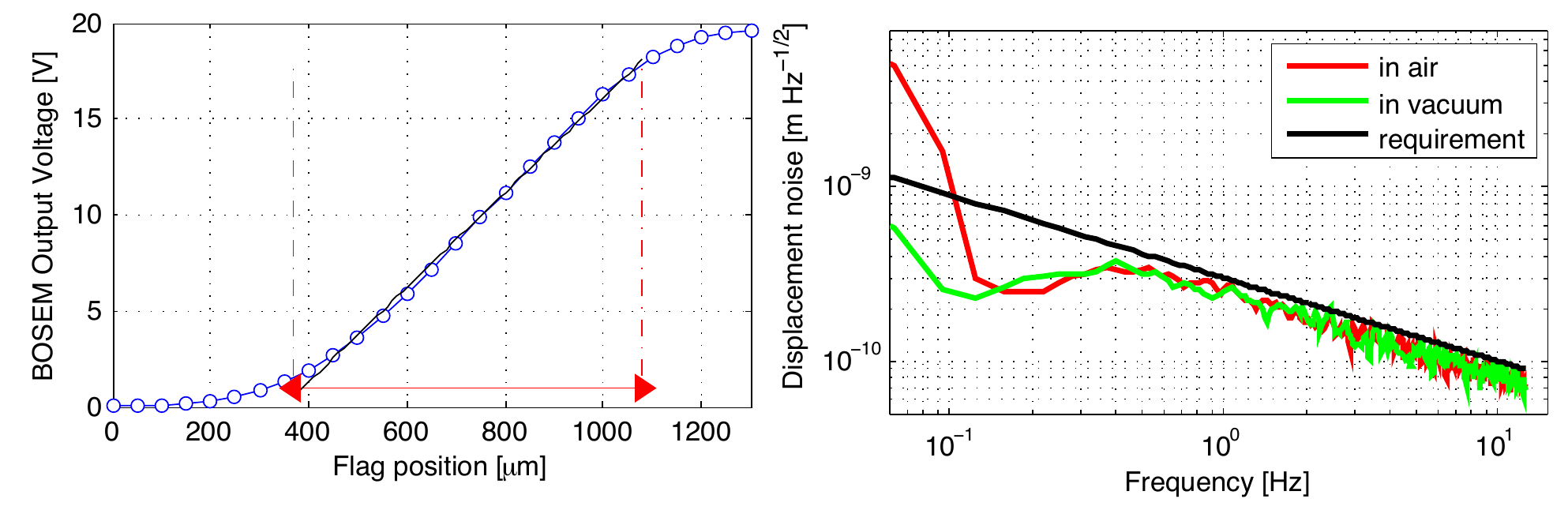}
\caption{\label{fig:BOSEM:perf} Typical performances of the BOSEMs displacement sensors. Left: response of BOSEM readout as function of position of the flag. Nominal measuring range (red lines), and fit in the linear region (black line) are also shown. Right: displacement sensitivity for a BOSEM unit, measured in-air and in-vacuum. The \aligo~requirements, nominally 3$\times 10^{-10}$\,m\,Hz$^{-1/2}$ at 1\,Hz and 1$\times 10^{-10}$\,m\,Hz$^{-1/2}$ at 10\,Hz \cite{strain:actuation:1}, are shown for comparison.}
\end{center}
\end{figure}

\begin{figure}[t]
\begin{center}
\includegraphics[width=13.7cm]{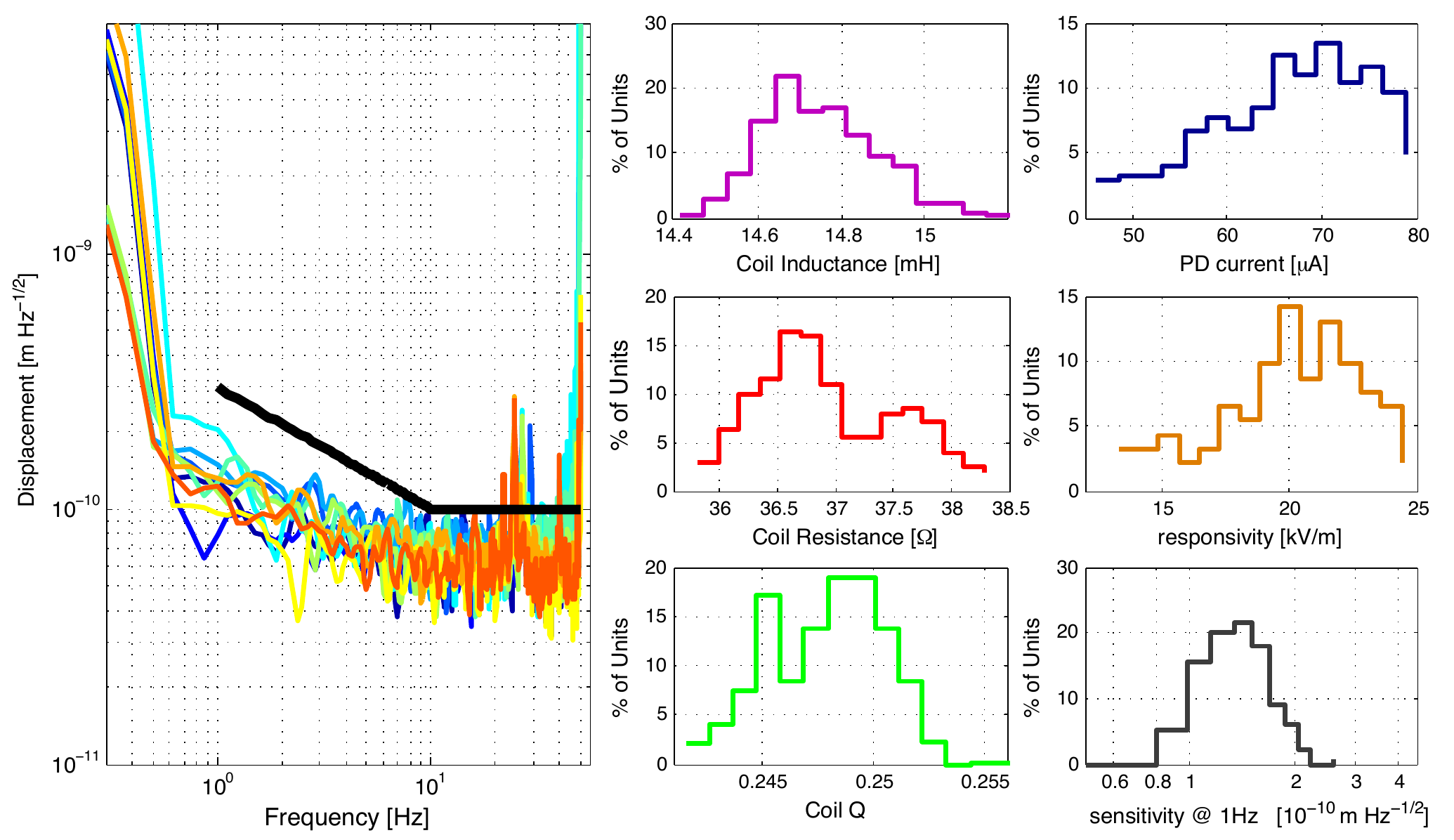}
\caption{\label{fig:BOSEM:prod} Typical performances of the BOSEM production articles. Left panel: displacement noise spectra of ten BOSEM samples, compared with the \aligo~requirement (black line). 
The spikes between 20\,Hz and 50\,Hz are mains-related pick-up and are artefacts of the measurement system. Central column: 
some statistics on the reproducibility of the BOSEMs electrical properties for $\sim600$ BOSEM articles: inductance (nominal design value 14.7\,mH), resistance (37.6\,$\Omega$) and electrical $Q$ ($243\times 10^{-3}$), to be compared with $\pm5\%$ tolerances from the requirements \cite{BOSEM}. 
Right column: distribution of the PD currents measured from $\sim600$ units (nominal design value 62.5\,$\mu$A, tolerance $\pm28\%$), and some statistics on responsivity and displacement noise at 1\,Hz measured for the fully characterised BOSEM units (about $\sim 20\%$ of the total). 
The large dispersion 
in the responsivity curve is mainly driven 
by irreproducibility of the 
off-the-shelf LEDs 
radiant intensities (up to 50\%), 
combined with a large dispersion in the quantum efficiencies of the PDs (as inferred from the PD half-light measurements shown here). However 
we remark 
that PD data refer to the entire 
production batch, while responsivities are measured 
only over 20\% of the population. 
}
\end{center}
\end{figure}

The optical read-out works following the shadow-sensing detection scheme depicted in Fig.\,\ref{fig:BOSEM:scheme}\,(right): an opaque object (`flag') is used to obscure a fraction of the light cast by a LED 
onto a 
single quadrant 
photodiode (PD), 
generating 
a current in the PD 
which is related 
to the position of the flag. 
A set of lenses and a mask are also included to improve the collimation of the beam emitted from the LED. This ensures that only paraxial rays hit the PD
sensing area, improving linearity and reducing noise. 
The flag is a 
25\,mm$\times$3\,mm {Al} cylinder, 
which moves along its longitudinal axis. 
The shape and size of the flag were chosen after investigating a range of possible 
geometries 
\cite{lockerbie:DCC:2004:2} 
\footnote[2]{More recent developments emerged during commissioning of the suspensions have suggested a small modification
in the design, with a reduced flag thickness and flattened sides 
along the non-measurement axis to further facilitating the installation process,
and a wider flag size in the orthogonal direction to decrease sensing cross-coupling 
\cite{flag:change}.}.
The BOSEMs use commercial infra-red opto-electronic 
components 
(Vishay TSTS7100 LEDs, BPX65 PDs) 
with emission and detection peaks around 950\,nm 
and negligible emission at the main interferometer wavelength (1064\,nm). 
The LED is operated at about 15\% of its maximum radiant power (10\,mW), 
which combined with the PD output (0.55\,A/W) 
results in 
a mean photocurrent at the PD output of $i_{\rm pd}\simeq 62.5 $\,$\mu$A, when the flag is in half-light position (i.e., PD sensing area obscured by 50\%).
The measuring range 
of the optical sensor is about 700\,$\mu$m, 
determined by the dimensions of the PD sensing area. The average responsivity, 
when measured with the accompanying satellite box amplifier, 
ranges around 
R$_{\rm esp}\sim$\,20\,kV\,m$^{-1}$, 
as shown in Fig.\,\ref{fig:BOSEM:perf}\,(left). 
The ultimate limit to the BOSEM read-out sensitivity is set by photo-current shot-noise in the PD, and it 
can be expressed as 
$S^{1/2}_x = \sqrt{2 \cdot e \cdot i_{\rm pd}}\cdot Z_R / {\rm R_{esp}}$, 
with $e$ is the electron charge and $Z_R=320$\,k$\Omega$ is the resistance of the trans-impedance amplifier at the output of the PD, resulting in an ultimate 
displacement sensitivity about 
$7\times 10^{-11}$\,m\,Hz$^{-1/2}$. 
The measured BOSEM sensitivity, as shown in Fig.\,\ref{fig:BOSEM:perf}\,(right) and Fig.\,\ref{fig:BOSEM:prod}\,(left) in comparison with the \aligo~requirement,
is typically limited by photo-current shot-noise in the PD at frequencies above 10\,Hz,
and by {\it 1/f} photo-current noise in the LED at lower frequencies.

To actuate on the suspended test masses, the BOSEM incorporates a magnet
which is rigidly mounted with the flag and it is concentrical to a coil purposely driven by the coil-drivers (see Fig.\,\ref{fig:BOSEM:scheme}).
To cater for the large weight
of the suspended masses, 
the BOSEMs provide peak actuation forces 
up to 
about hundred mN per actuator, 
about one order of magnitude 
enhancement with respect to the Initial LIGO OSEM. 
This is achieved with 
a modified coil design (800 turns of polyimide coated 32-AWG wire) 
and with 10\,mm$\times$10\,mm Nd-B-Fe 
cylindrical magnets (magnetic moment $\mu \sim 0.5$\,Am$^2$) \footnote[3]{Due to the relaxed actuation strength requirements, on intermediate and penultimate stages the BOSEMs use SmCo magnets and different size and magnetic moments \cite{robertson:magnets}.}
\cite{strain:actuation:1,strain:actuation:2,robertson:magnets}. Residual magnetic forces are kept below 5mN, by use of 
low magnetic susceptibility Al6082 for the 
mechanical parts, and with a proper separation of actuator and sensor parts. 
Furthermore, the BOSEM magnets are assembled in quadrupolar configuration on the test masses in order to null the resulting overall magnetic moment and thus minimise their susceptibility to external magnetic fields.
The electromagnetic cross-talk from the actuation coil to the readout channel was measured to be below 
$6\times 10^{-5}$\,m/A
and it was found to be most prominent only at frequencies above 30 Hz, i.e. where this 
coupling is naturally mitigated by the mechanical transfer function of the quadruple pendulum , and to be 
arising from non-shielded harnesses of the test equipment to and from the Satellite Boxes (however shielding is foreseen in the final design).

The BOSEM design incorporates several other functional features 
which have been included to facilitate large scale production and reproducibility of the device's performance, and to improve their usability. 
The overall compact design (less than $40\times40\times 40$\,mm$^3$ volume and $\sim$170\,g mass) incorporates 
a user-friendly mounting and alignment mechanism, which allows 10\,mm longitudinal and 1.5\,mm lateral adjustments 
with respect to the flag position, during installation process.
Performance reproducibility is improved also with the use of standard-leaded 
components, in place of the 
originally glued 
surface-mount ones,
which are press-fit into ceramic sleeves, encapsulated with retainers 
in aluminium alloy assemblies, and 
eventually screwed in to position,
minimising the overall misalignment of the components. 
The absence of potentially outgassing adhesives 
from the fabrication and alignment process 
provides also an enhanced UHV compatibility of the assembly. 
Finally, the BOSEMs use flexi-circuit for interconnections to/from LED, PD and the coil, 
and miniature D-type 
connectors, to facilitate assembly/disassembly and testing of individual components.
 
Production of the BOSEMs was conducted in Birmingham's clean room facilities, 
after fabrication of mechanical parts by external contractors, followed by in-house inspection against tolerances, 
and subsequent cleaning and baking according to \aligo~UHV requirements. 
All opto-electronic components have been pre-screened to identify infant mortality of the devices and to ensure performance consistency. 
A more intensive pre-screening of the 
LEDs was necessary to avoid an observed 
non-reproducibility in the noise performances of the LEDs themselves, degrading the BOSEMs displacement sensitivity by about a factor two above specifications.
This was eventually related to a non-reproducible deposition 
of the adhesive used in the LED package sealing during the manufacturing process \cite{aston:PHD:2011}
which was 
therefore
impossible to identify 
via non-destructive physical inspection of the devices, 
and required extensive noise-characterisation of each individual LED unit, and the selection of 
only the compliant ones (about 75\% of the total).

All assembled BOSEMs have been tested using the Automated Test Equipment - ATE,
also part of 
the deliverables~\footnote[1]{The BOSEM-ATE consists of an LCR Meter, Digital Multimeter, USB interface box, and software to run the tests and store the data on a PC.} - 
to ensure consistency across the entire batch, 
reject non-compliant units 
and to provide a fully trackable record of the relevant 
properties (coil resistance, inductance and electrical $Q$, PD current).
About 20\% of all the delivered BOSEM units underwent full end-to-end characterisation test using 
\aligo~representative equipment, providing a complete record of 
operating range, responsivity, and displacement sensitivity \cite{coildrivers}. 
Examples of the results from this characterisation campaign, and some statistics on the reproducibility of the production articles performances 
are shown in Fig.\,\ref{fig:BOSEM:prod}.

\subsection{Satellite boxes \label{Sec:SatBox}}
\begin{figure}[t]
\begin{center}
\includegraphics[width=12.3cm]{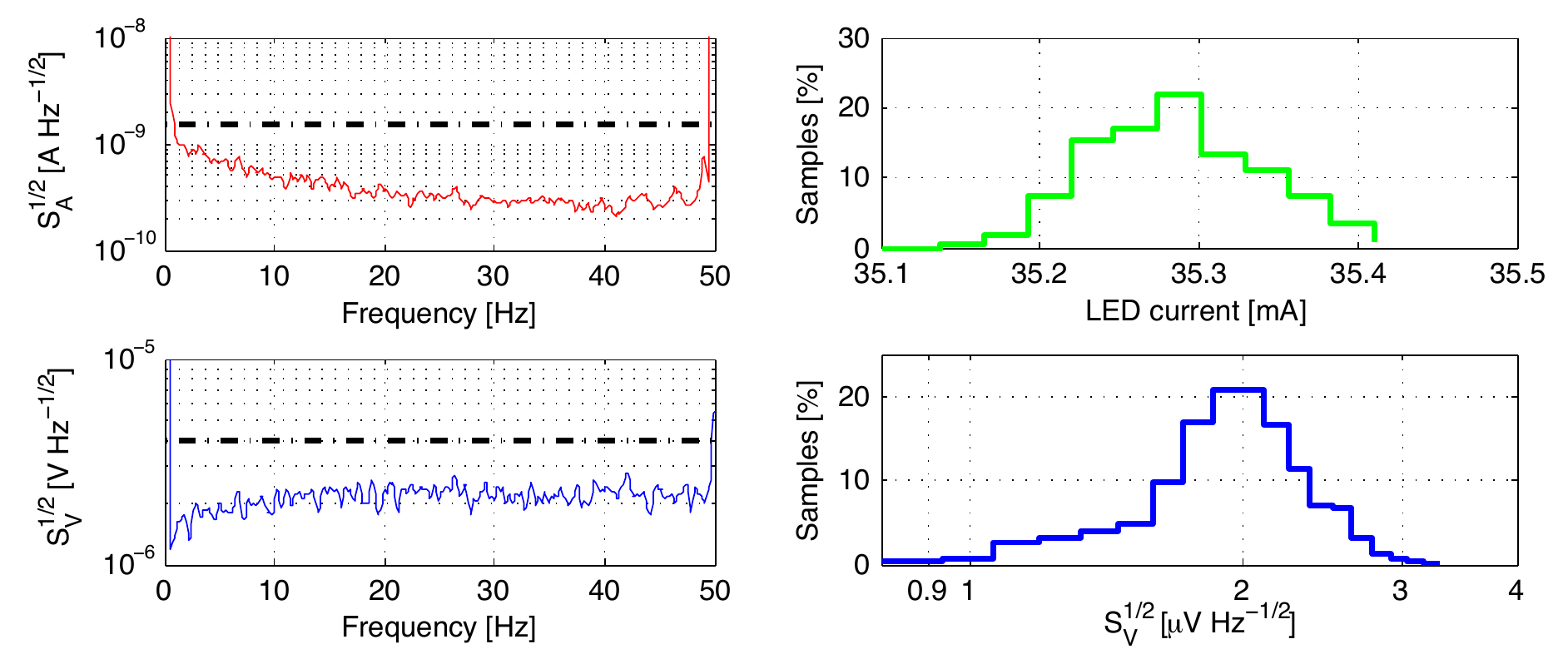} 
\caption{\label{fig:SBnoise} Satellite boxes typical performances. 
Left, LED current supply (top) and PD readout voltage noise performances (bottom) compared to requirements (black dashed lines). 
Right, distribution of the LED current (top, tolerance $\pm5\%$) and PD amplifier voltage noise at 10\,Hz (bottom) measured over $\sim230$ Satellite box units ($\sim 920$ channels).
}
\end{center}
\end{figure}
The `satellite boxes' 
incorporate a current supply for the BOSEM LEDs, a voltage amplifier for the BOSEM PD signal, and provide electrical pass-through from actuation electronic units to the coils.
Monitoring circuits are also included in the unit to monitor its activity and status. Each satellite box is equipped with four 
current sources and 
amplifier channels, serving four OSEMs units at a time. 

The current source generates a stable, low noise current  
with nominally 35\,mA and a measured noise level of $\sim0.5$\,nA\,Hz${^{-1/2}}$ at 10\,Hz, 
which corresponds to an equivalent displacement noise of $\sim 3\times10^{-11}$\,m\,Hz$^{-1/2}$, 
three times lower than the requirement 
\cite{carbone:DCC:2010}. 
The PD amplifier converts the PD current output into a -/+10\,V signal, 
with intrinsic voltage noise not exceeding the 4\,$\mu$V\,Hz$^{-1/2}$ upper limit requirement \cite{ref:SB} 
and equivalent to a displacement noise  of $5\times10^{-11}$\,m\,Hz$^{-1/2}$  at 10\,Hz. 

Satellite box manufacturing was conducted by outsourcing the production of PCBs to external contractors, and consequently populating and assembling the units in-house. 
Functional and performance testing has been performed on all the production articles 
with a dedicated 
ATE, and recorded for performance traceability \cite{coildrivers}. 
The characteristic LED current noise, the PD amplifier voltage noise and some statistics on reproducibility of the performances of the satellite box production units are presented in Fig.\,\ref{fig:SBnoise}.

\subsection{Coil-driver units \label{Sec:CoilD}}
The `coil-driver' units \cite{coildrivers} amplify the signal from the digital control system to drive 
the BOSEMs coils for actuating on the suspension stages. 
Each coil-driver unit contains four current drive circuits, 
and is equipped with 
read-back circuitry for remote 
monitoring of the unit performance.
\begin{table}[b]
\caption{\label{coil:drivers} Performance requirements for the quadruple suspension coil-driver units 
\cite{strain:actuation:1,strain:actuation:2,heefner:DCC:2006,barton:2010}.
}
\begin{center}
{\small \begin{tabular}{ c || c c c }
Coil-driver & Q-TOP & UIM & PUM\\
\hline \hline \hline

dynamic range & 
\begin{tabular}{c} $\pm$\,200mA\\ (continuous)\end{tabular} & {\footnotesize \begin{tabular}{c} 2\,mA$_{\rm rms} <$ 1\,Hz\\ 16\,$\mu$A$_{\rm rms}$   @ 100\,Hz \end{tabular}}& 
{\footnotesize \begin{tabular}{c} 16\,mA$_{\rm rms}$ \\ (200 - 5000\,Hz)\end{tabular}} \\

\hline

noise @ 1\,Hz & 1\,nA\,Hz$^{-1/2}$  		& 0.5\,nA\,Hz$^{-1/2}$	& 20\,nA\,Hz$^{-1/2}$\\
noise @ 10\,Hz         & 73\,pA\,Hz$^{-1/2}$  		& 3\,pA\,Hz$^{-1/2}$ 		& 4\,pA\,Hz$^{-1/2}$\\
noise @ 100\,Hz & 1000\,nA\,Hz$^{-1/2}$  	& 200\,nA\,Hz$^{-1/2}$ 	& 5\,nA\,Hz$^{-1/2}$\\
noise @ 1000\,Hz & 1000\,nA\,Hz$^{-1/2}$  	& 1000\,nA\,Hz$^{-1/2}$ 	& 1000\,nA\,Hz$^{-1/2}$\\
\hline
actuation strength & 1.7\,N\,A$^{-1}$& 1.7\,N\,A$^{-1}$ & 0.03\,N\,A$^{-1}$\\
\hline
max force noise @ 10\,Hz & 40\,pN\,Hz$^{-1/2}$ & 8\,pN\,Hz$^{-1/2}$ &  0.1\,pN\,Hz$^{-1/2}$ \\
\hline
\end{tabular}
}
\end{center}
\label{default}
\end{table}
Within the \aligo~suspensions, 
the coil-driver units serve two functions: first, to provide strong
actuation authority for the damping of the motion of the suspended masses, to allow for acquisition 
of the interferometer system; second, to provide quiet actuation to allow for control of the overall interferometer during detector science mode.

Six different types of coil-drivers have been developed, each designed for
the specific performance, in terms of frequency response, output current and noise levels, 
each stage and type of the suspension required. The coil-drivers actuation functions are mainly three: 
`top' coil-drivers 
serve for the static alignment/positioning of the optics 
and correction of slow drifts, 
and more generally 
compensate for the very low frequency motion of the optics below 1\,Hz; 
`intermediate' coil-drivers 
control the low frequency suspension's dynamics in the frequency range between 1\,Hz and 10\,Hz; 
`lower' coil-drivers 
actuate at high frequencies 
(above 10\,Hz) and serve for interferometer lock acquisition (`acquisition mode') and lock maintenance (`run mode'). 
The performance requirements for each of the coil-driver units employed in the quadruple suspensions control system,
named Quadruple Top (Q-TOP), Upper Intermediate (UIM) and Penultimate (PUM) 
is given in Table\,\ref{coil:drivers} \cite{heefner:DCC:2006}. 
Coil-drivers units 
designed for 
shorter suspension types have similar frequency dependence 
and slightly relaxed noise requirements 
\cite{heefner:DCC:2006}.

\begin{figure}[t]
\begin{center}
\includegraphics[width=13.6cm]{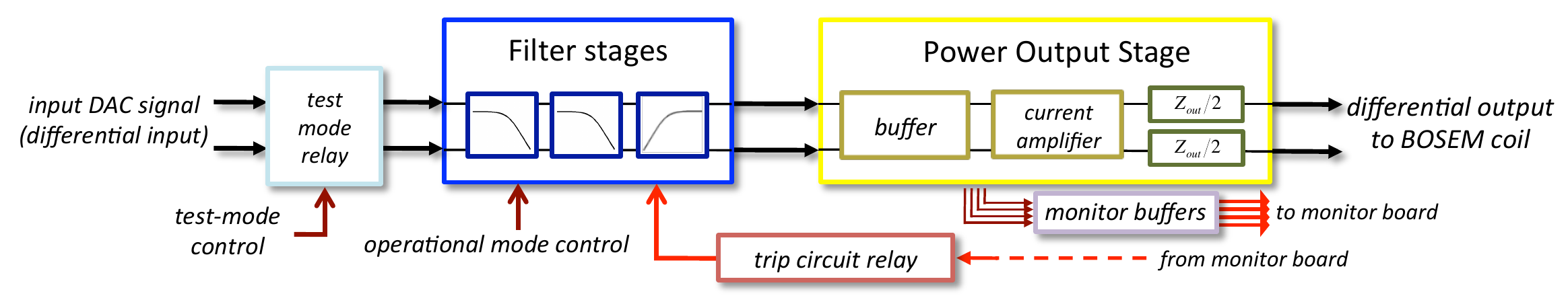}
\caption{\label{fig:CD:scheme} 
Schematic of the coil drivers working principle.
To minimise coupling to electrical ground noise, the input to the unit from the  DAC is differential and balanced with respect to ground.
The electrical signals are then kept independent and symmetrical with respect to 
ground in the unit by using two independent chains of amplifiers, preventing ground noise being introduced within the unit.
The ÔtestÕ mode  and  operational modes can be selected remotely by the user.
Different numbers of low pass and/or high pass filters are implemented in the various units, depending on the specific type, and filters may be enabled/disabled within the unit according to the required operational mode.
Signals from the power output stage are sent to the monitor boards for diagnostics, before driving the BOSEM coils.
In units with high current modes, based on the monitor board signals, a trip relay can disable the driver unit 
preventing overheating and damage of the 
coils which would occur if excessive currents are supplied to the coil for prolonged periods.}
\end{center}
\end{figure}

\begin{figure}[t]
\begin{center}
\includegraphics[width=13.1cm]{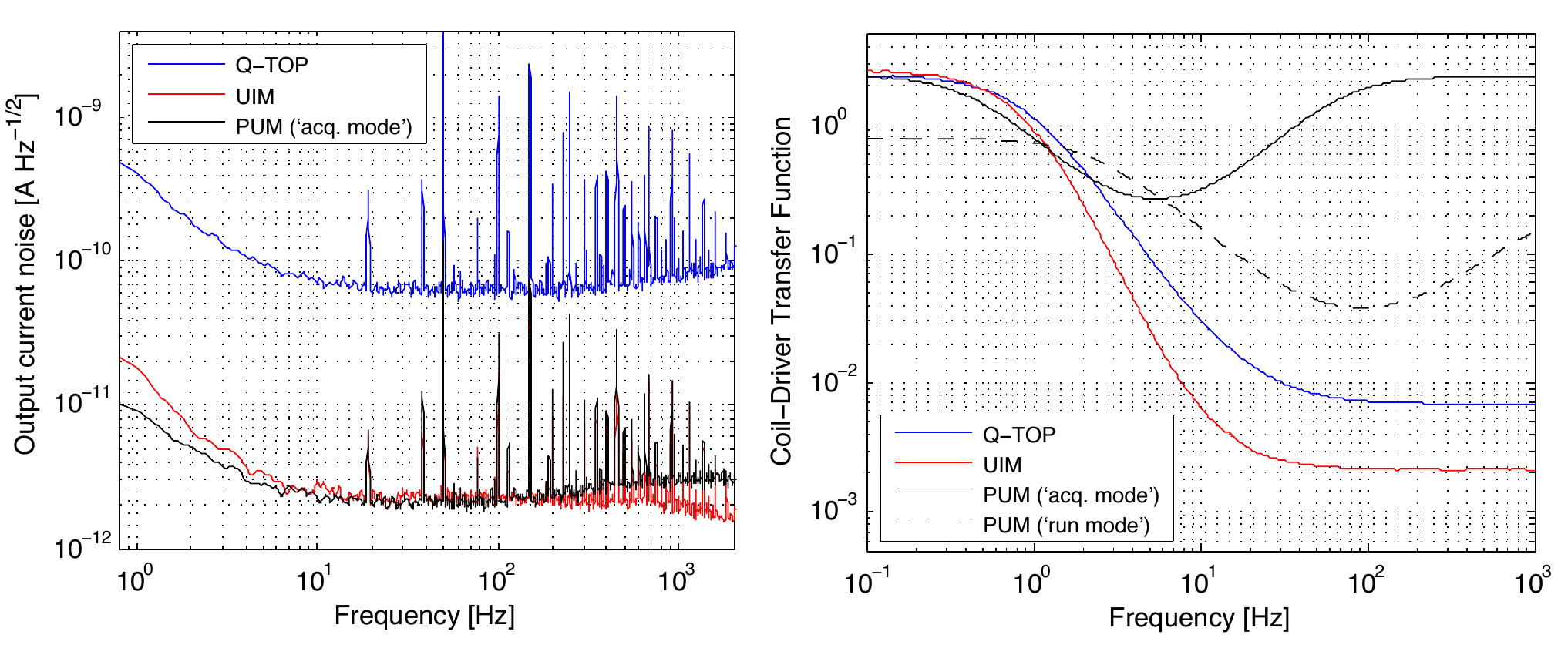}
\caption{\label{fig:CD:perf} Measurements of the performances for the 
coil-driver units used in the quadruple suspensions (Q-TOP, UIM, PUM). 
Left: output current noise, as measured when grounding the coil-driver inputs. The spikes present in all three noise measurement curves are mains-related pick-up, which are artefacts of the measurement system.
Right: coil-drivers frequency responses. PUMs are here shown in both operational modes, `acquisition' and `run' mode.}
\end{center}
\end{figure}

The coil-drivers are designed to generate low noise, high drive currents based on the input signals 
of up to 20\,V$_{\rm pp}$ and with a maximum input
noise of $S^{1/2}_{V_{in}}\approx 100$\,nV\,Hz$^{-1/2}$,
coming from the digital control system. 
A sketch of the coil driver's working principle is shown in Fig.\,\ref{fig:CD:scheme}.
The large dynamic range of the output current (up to $\sim$\,180\,dB as shown in Table\,\ref{coil:drivers}) is necessary to combine the diverse 
functionalities required for the BOSEM coil actuation system discussed above.
To provide the demanded high current at the output, the input signals are buffered with high-power 
operational amplifiers. 
The maximum values for the 
output current are set using suitable 
output impedances $Z_{\rm out}$ 
in series with the BOSEM coil resistance $R_{\rm Bosem}$. 
The output current noise is determined by the input voltage noise $S^{1/2}_{V_{\rm in}}$ from the digital control system and
by the intrinsic voltage noise $S^{1/2}_{\rm Int}$ generated inside the coil-driver circuitry, and it can be expressed as
$ S^{\frac{1}{2}}_{\rm out}=\left(S_{V_{\rm in} } \cdot |F|^2 + S_{\rm Int} \right)^{\frac{1}{2}}/\left(Z_{\rm out}+R_{\rm Bosem}\right)
$,
with $F$ the coil-driver transfer function. 
On one hand, $S^{1/2}_{\rm Int}$ is kept low with purposeful design of the circuitry itself, 
careful selection of the 
active components and suitable rejection of external noise, mainly due to electrical ground instability.
On the other, the different filtering stages included in the units provide the requested transfer functions 
to reduce the role of the input noise $S^{1/2}_{V_{\rm in}}$
down to a negligible level. 
Measurements of the typical performance of the coil-driver units used in the quadruple suspensions are presented in Fig.\,\ref{fig:CD:perf}, and show how these units agree with specifications over the entire bandwidth.  

As well as meeting the technical specifications 
discussed above, the coil-drivers were designed to comply with several additional operational requirements. 
Besides having a robust and reliable design that could 
guarantee 
a long operational life-cycle, 
a relatively large adaptability of the units was specifically requested, in particular to permit for 
later optimisations 
to the frequency responses of 
the coil-driver units 
as the understanding of the overall system is 
expected to
develop with the commissioning and operation of \aligo.
This was implemented by making the filter sections switchable 
by means of relays, 
chosen 
for their inherently lower noise.
Similarly, the different operational modes, e.g., `acquisition' or 'run' mode of the lower stages coil-driver units, are implemented with
relays switching between different output resistors and filters.
Furthermore, to provide 
serviceability and ease of modification
the electronic layout was intentionally non-extensively miniaturised and the PCBs were 
populated at low density,
to facilitate a rapid replacement 
of a potentially failing component, or to enable components to be added or their values changed.
\begin{figure}[b]
\begin{center}
\includegraphics[width=13.6cm]{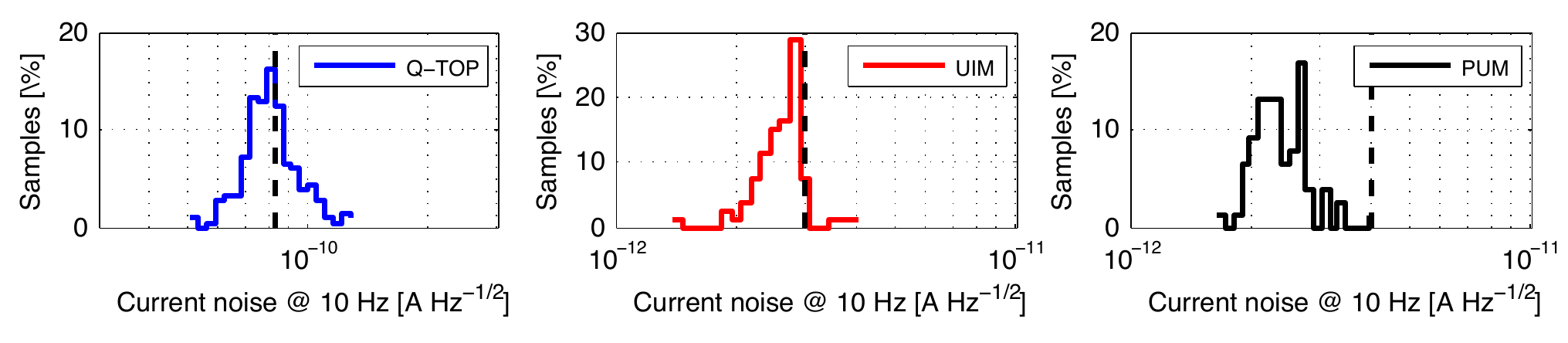}
\caption{\label{fig:CD:prod} 
Measurements of output current noise at 10\,Hz for the three coil-driver types used in the quadruple suspensions,
on a total of 80, 80 and 270 output channels respectively. 
The requirements are reported in each graph with black vertical dashed lines.
It is important to note that a fraction of the Q-TOP drivers were found to slightly exceed the current noise specifications at 10\,Hz ($\sim30\%$ of units above requirement),
however it was agreed to tolerate this small excess and rather relax the requirement 
to facilitate respecting the production schedule. 
}
\end{center}
\end{figure}

The coil-driver units represent a large proportion of the articles produced at the University of Birmingham for the \aligo~suspensions, for a total of about 270 units. 
As for the satellite boxes, coil-driver PCBs were designed in-house, manufactured by external contractors, and then assembled, tested and characterised 
in Birmingham. 
For all the coil-driver production units, all performance and functional requirements have been individually tested, validated and the results recorded for future reference \cite{coildrivers}.
Statistics on some of the output current noise performance of the quadruple suspensions units are shown
in Fig.\,\ref{fig:CD:prod} in comparison with the respective requirement.

\section{Conclusion}
\begin{figure}[t]
\begin{center}
\includegraphics[width=13.3cm]{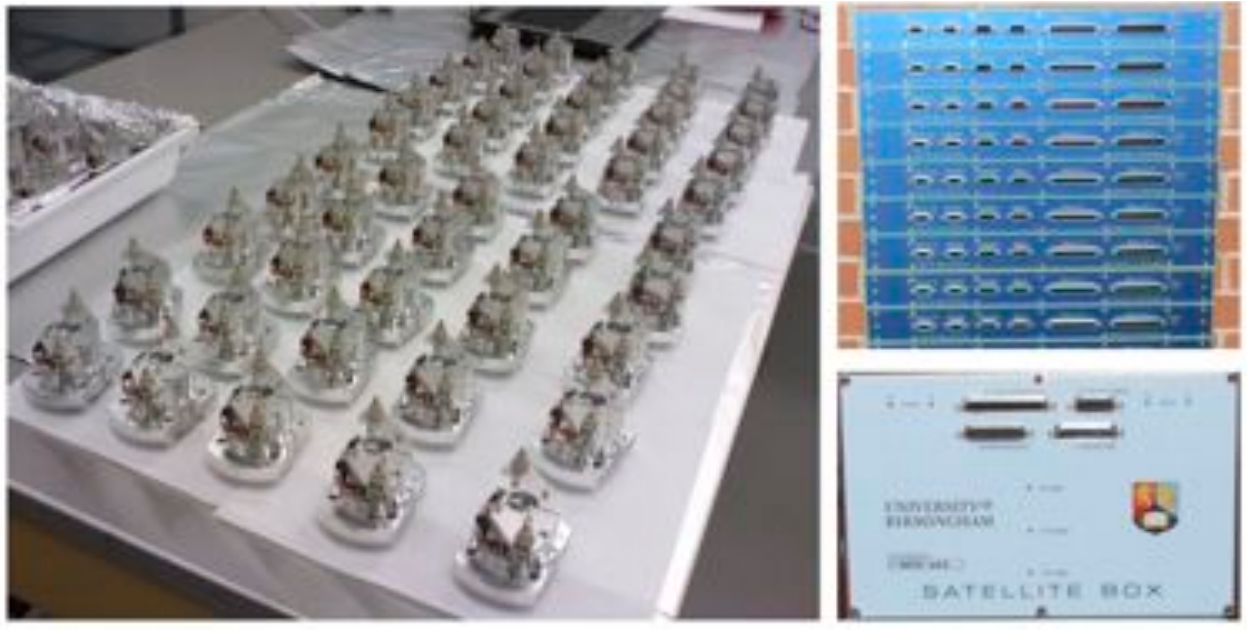}
\caption{\label{fig:deliverables} 
Pictures of the final production articles. 
On the left, BOSEM units during assembly and inspection in the 
clean room facilities. Top right: a 
stack of PUM coil-driver units. Bottom right: the front panel of  a satellite box.}
\end{center}
\end{figure}

Production and testing 
of the instrumentation 
discussed in the previous sections, 
and here shown in the photographs 
in Fig.\,\ref{fig:deliverables}, 
was completed by the end of the \aligo~UK project grant, which formally terminated in March 2011.
All the produced instruments were successfully characterised
and their performance validated to meet the stringent requirements for the achievement of the seismic isolation goals of the \aligo~suspensions. All items were shipped overseas to their destination at the \aligo~detector sites (1/3 of the production units to the Livingston Laboratory, 2/3 to Hanford) as well as to the prototypes. 
The instruments are currently being installed and commissioned on the different interferometer suspensions systems, and preliminary performance of the 
system 
is 
successful and 
encouraging for the future sensitivity of \aligo.

The sensing and actuation technology 
presented in this paper has been successfully pushed very far in all aspects of the required performance,
and further improvements are likely to demand substantial re-design effort.
With this aim, further R\&D work is underway 
to investigate improved solutions for sensing and actuation instrumentation 
for a potential employment in future GW detectors of the third generation. 

\ack

The authors gratefully acknowledge the Science and Technology Facilities Council (STFC) 
for funding of this project. We are grateful to the STFC, the University of Birmingham, 
the University of Glasgow and the University of Strathclyde for financial support in the United Kingdom  
and to the United States National Science Foundation (NSF) for the construction and operation of the
LIGO Laboratory and for US funding for Advanced LIGO. 
NAR and JH are supported by NSF grant PHY - 0107417. 
This document has been assigned LIGO Document Number LIGO-P1100208.


\section*{References}
\begin{thebibliography}{9}
\bibitem{LIGO} B.P. Abbott et al 2009 Rep. Prog. Phys. 72 076901, http://www.ligo.org/
\bibitem{GEO} S. Hild et al, Class. Quantum Grav., 23, S643, 2006, http://www.geo600.org
\bibitem{VIRGO} F. Acernese et al, Class. Quantum Grav., 25, 2008, http://www.ego-gw.it/ 
\bibitem{harry:CQG:2009} G.M. Harry and the LIGO Scientific Collaboration, Class. Quantum Grav. 27 084006, 2010. 
\bibitem{geohf} B. Willke et al, Class. Quantum Grav., 23, S207, 2006 
\bibitem{avirgo} T. Accadia et al, Class. Quantum Grav. 28, 114002, 2011 
\bibitem{ligo:seis} R. Abbott et al, Class. Quantum Grav. 21 S915, 2004
\bibitem{robertson:CQG:2002} N. Robertson et al, Class. Quantum Grav. 19 (2002) 4043-4058
\bibitem{ref:monolithics} A. Cumming et al, Class. Quantum Grav., 26, 21, 215012 (2009) and references therein
\bibitem{ALUK:proposal} J. Greenhalgh, {{\it Advanced LIGO UK Participation}, LIGO-G030424-00-K,} 2003 
\bibitem{fritschel:DCC:1999} P. Fritschel.	{\it Characterization and comparison of a potential new  local sensor}, LIGO-T990089-00 1999.
\bibitem{AOSEM} R. Abbott et al, {\it Advanced LIGO OSEM Final Design Document}, LIGO-T0900286-v2, 2009
\bibitem{elec:drive} N.A. Lockerbie, {\it Electrostatic Driver}, LIGO-T1100231-v2, 2011
\bibitem{shapiro:2012} B. Shapiro et al, {\it Modal Damping of a Quadruple Pendulum for Advanced Gravitational Wave Detectors}, 2012 American Controls Conference 
\bibitem{strain:2012} K. A. Strain et al, Rev. Sci. Instrum. 83, 044501 (2012)
\bibitem{virgoSA} V. Dattilo for the VIRGO collaboration, Physics Letters A 318 (2003) 192 
\bibitem{virgoLS} F. Acernese et al, Astroparticle Physics 20 (2004) 617 
\bibitem{speake:2005} C.C. Speake et al, Class. Quantum Grav. 22 S296 (2005) 
\bibitem{shoemaker} D. Shoemaker et al, Physical Review D 38, 2, 423 (1988)
\bibitem{romie:DCC:2003} J. Romie, 
{{\it Hybrid OSEM assembly specification}, LIGO-E030084-02-D}, 2003 
\bibitem{strain:DCC:2004:feb} K.A. Strain et al, 
{{\it Recommendation of a design for OSEM sensors}, LIGO- E040108-01-K}, 2004 
\bibitem{strain:DCC:2004} K.A. Strain, 
{{\it Input to the OSEM selection review decision}, LIGO- T040110-01-K}, 2004. 
\bibitem{lockerbie:DCC:2004:1} N.A.~Lockerbie, 
{{\it Measurement of LIGO hybrid OSEM sensitivity} LIGO-T040106-01-K}, 2004. 
\bibitem{lockerbie:DCC:2004:2} N.A.~Lockerbie, 
{{\it Measurement of shadow-sensor displacement sensitivities} LIGO-T040136-00-K}, 2004. 
\bibitem{flag:change} M. Evans, 
{{\it BOSEM Flat Magnet Flag, aLIGO SUS}, LIGO-D1100573-v5}, 2011 
\bibitem{strain:actuation:1} K.A. Strain, 
{{\it Advanced LIGO suspension: summary noise calculations as input to the UK electronics PDR}. LIGO-T050122-00-K}, 2005 
\bibitem{strain:actuation:2} K.A. Strain, 
{{\it Increased strength Advanced LIGO ITM/ETM suspension PM and UIM Actuators}, LIGO-T060001-00-K}, 2006 
\bibitem{robertson:magnets} N. Robertson et al, {\it Magnet sizes and types and OSEM types in Adv. LIGO suspensions}, LIGO- M0900034-v4, 2011
\bibitem{BOSEM} S.M. Aston, 
{{\it BOSEM Design Document \& Test Report}}, LIGO-T050111-04-K, 2009
\bibitem{aston:PHD:2011} S.M. Aston, {\it Optical Read-out Techniques for the Control of Test-masses in Gravitational Wave Observatories}, PhD Thesis, University of Birmingham, 2011.
\bibitem{coildrivers} L. Carbone et al, 
{{\it Summary of technical documentation for University of Birmingham deliverables}}, LIGO-T1100185-v2, 2011, and references therein, 
\bibitem{carbone:DCC:2010} L. Carbone et al, 
{{\it Satellite Boxes Current-Source noise characterisation}}, LIGO-T1000630-v2, 2010. 
\bibitem{ref:SB} D. Hoyland et al,
{\it OSEM Drive Electronics Interface Control Document}, LIGO-E040374-00-K, 2004 
\bibitem{heefner:DCC:2006} J. Heefner, {\it AdL Quad Suspension UK Coil-Driver Design Requirements}, LIGO-T060067-00-C (2006);
J. Heefner, {\it  AdL Beam Splitter, Input Mode Cleaner, Large Recycling and Small Recycling Triple Suspension Electronics Requirements}, LIGO-T080065-E1-C (2008);
info also available at \href{http://www.its.caltech.edu/~rana/aLIGO/suselecreq.html}{www.its.caltech.edu/$\sim$rana/aLIGO/suselecreq.html} 
\bibitem{barton:2010} M. Barton, 
{ {\it Calculation and measurement of the OSEM actuator sweet spot position}, LIGO-T1000164-v3,} 2010
\end{thebibliography}
\end{document}